\def\ube13{UBe$\rm_{13}$}
\def\prinag{PrInAg$\rm_2$}

\documentstyle[aps,prb,twocolumn,epsf]{revtex}
\begin{document} 
\draft

\def\dfrac#1#2{{\displaystyle{#1\over#2}}}
\twocolumn[\hsize\textwidth\columnwidth\hsize\csname @twocolumnfalse\endcsname
to be published in PRB Rapid Comm. \hfill{LA-UR-98-2244}

\title{Magnetic Field Dependence of the Low-Temperature Specific Heat in PrInAg$\rm_2$: Support for a Non-Magnetic Heavy-Fermion Ground State}

\author{R. Movshovich,$^1$  A. Yatskar,$^2$ M. F. Hundley, $^1$ P. C. Canfield,$^3$ and W. P. Beyermann$^2$ }
\address{$^1$Los Alamos National Laboratory, Los Alamos, New Mexico 87545 \\ $^2$University of California, Riverside, California 92521\\ $^3$Ames Laboratory and Iowa State University, Ames, Iowa 50011}

\date{\today}

\maketitle

\begin{abstract} 

In order to elucidate the nature of the ground state in the heavy-electron system PrInAg$\rm_2$, we measured its specific heat for temperatures between 60 mK and 2 K in magnetic fields up to 9 T. The peak maximum of the low temperature anomaly in the specific heat (interpreted as Kondo anomaly) shifts from 0.42 K at zero field to 0.45 K at 6 T, and to 0.5 K at 9 T. The data at 3 T (after subtracting the low temperature tail due to a Schottky anomaly) is practically indistinguishable from the zero field data. A low temperature nuclear Schottky anomaly from a  hyperfine enhancement of the Pr nuclei was observed in field. Both enhanced hyperfine interaction and the insensitivity of the specific heat anomaly to fields support the hypothesis that the Kondo effect in this system has a non-magnetic origin. 

\end{abstract}


]
\narrowtext

Recently, a novel electronic ground state was proposed for PrInAg$\rm_2$,~\cite{yatskar96} which
is characterized by a very large value of $\sim 7$ $\rm J/mol\ K^2$ for the Sommerfeld coefficient $\gamma = C/T$ for the specific heat. It was suggested that quadrupolar fluctuations of the Pr ions and their interaction with the charge of the conduction electrons lead to a heavy-fermion ground state, in direct analogy to the magnetic Kondo interaction between an impurity spin and that of a conduction electron. The identification of the non-magnetic nature of the ground state was based on measurements of several properties that determined the arrangement of the crystal-electric field (CEF) levels in this system. Inelastic neutron scattering experiments imply that the ground state is a non-magnetic, non-Kramers doublet ($\Gamma_3$) with a magnetic triplet ($\Gamma_4$) as the first excited state.~\cite{galera84,yatskar96} The low temperature susceptibility appears to saturate below 20 K to the value of 0.04 emu/mol, before upturning below 7 K. Calculation of the Van Vleck contribution from excited crystal field levels determined from neutron scattering yields 0.043 emu/mol, close to the value suggested by the data between 7 and 20 K. On the basis of this agreement it was suggested that susceptibility is of Van Vleck character.\cite{yatskar96}
This interpretation is also supported by the magnetic contribution to the specific heat, which is well described by a doubly degenerate ground state with two triply degenerate and one singly degenerate excited states.~\cite{yatskar96} These measurements support identification of the ground state of Pr ions in PrInAg$\rm_2$ as a non-magnetic non-Kramers doublet $\Gamma_3$.

A non-magnetic $\Gamma_3$ doublet ground state has  been invoked in the past to explain some of the thermodynamic and neutron scattering  properties of the heavy-fermion system UBe$\rm_{13}.$~\cite{cox87}  In particular, the low-temperature specific heat of UBe$\rm_{13}$ is nearly field independent, with $\gamma$ changing by few percent at 11 T.~\cite{stewart84} According to Hund's rules both $5f^2$ U$\rm^{4+}$ and $4f^2$ Pr$\rm^{3+}$ have a total angular momentum of $J = 4$. Since the U ions in UBe$\rm_{13}$ and Pr ions in PrInAg$\rm_2$ are at points of cubic symmetry, the CEF environments are identical within the point charge model. If indeed the ground state of Pr is a non-magnetic $\Gamma_3$ doublet, the effect of the magnetic field on the heat capacity of PrInAg$\rm_2$ should be small, just as it is for UBe$\rm_{13}$. 

In this article we present the specific heat data for PrInAg$\rm_2$ in magnetic fields up to 9 T, which provides a strong test of the hypothesis of a field-independent Kondo effect involving a non-magnetic, non-Kramers $\Gamma_3$ doublet  ground state in PrInAg$\rm_2$. We find that the low-temperature specific heat in a magnetic field consists of two contributions: a largely field-independent anomaly with a peak between 0.4 and 0.5 K, and a low temperature tail with a magnitude that increases with field. The weak field dependence of the first contribution (which is the subject of previous work at zero field~\cite{yatskar96}) proves its non-magnetic origin. The low temperature tail results from the hyperfine enhanced nuclear Schottky anomaly of the Pr ions in the non-magnetic electronic ground state. These results give additional support to the proposed quadrupolar Kondo effect in \prinag. 

The samples were grown by melting stoichiometric amounts of Pr, Ag, and In in a Ta crucible (details of sample preparation and characterization are described in Ref.~\onlinecite{yatskar96}). X-ray diffraction measurements were consistent with the previously reported cubic Heusler crystallographic structure of PrInAg$\rm_2$.~\cite{galera84} A residual resistivity ratio between 300 K and 4 K of $\sim 6.5$, and a residual resistivity  of $\rho_0 \approx 5.5$ $\mu\Omega$cm demonstrate the high quality of the samples. Specific heat data were collected with a quasiadiabatic technique,~\cite{quas_adb} where ruthenium oxide thick film resistors~\cite{sota} were used for thermometry. These resistors were previously calibrated as a function of temperature in a magnetic field against a thermometer placed in a 

\begin{figure}
\epsfxsize=3.5in
\centerline{\epsfbox{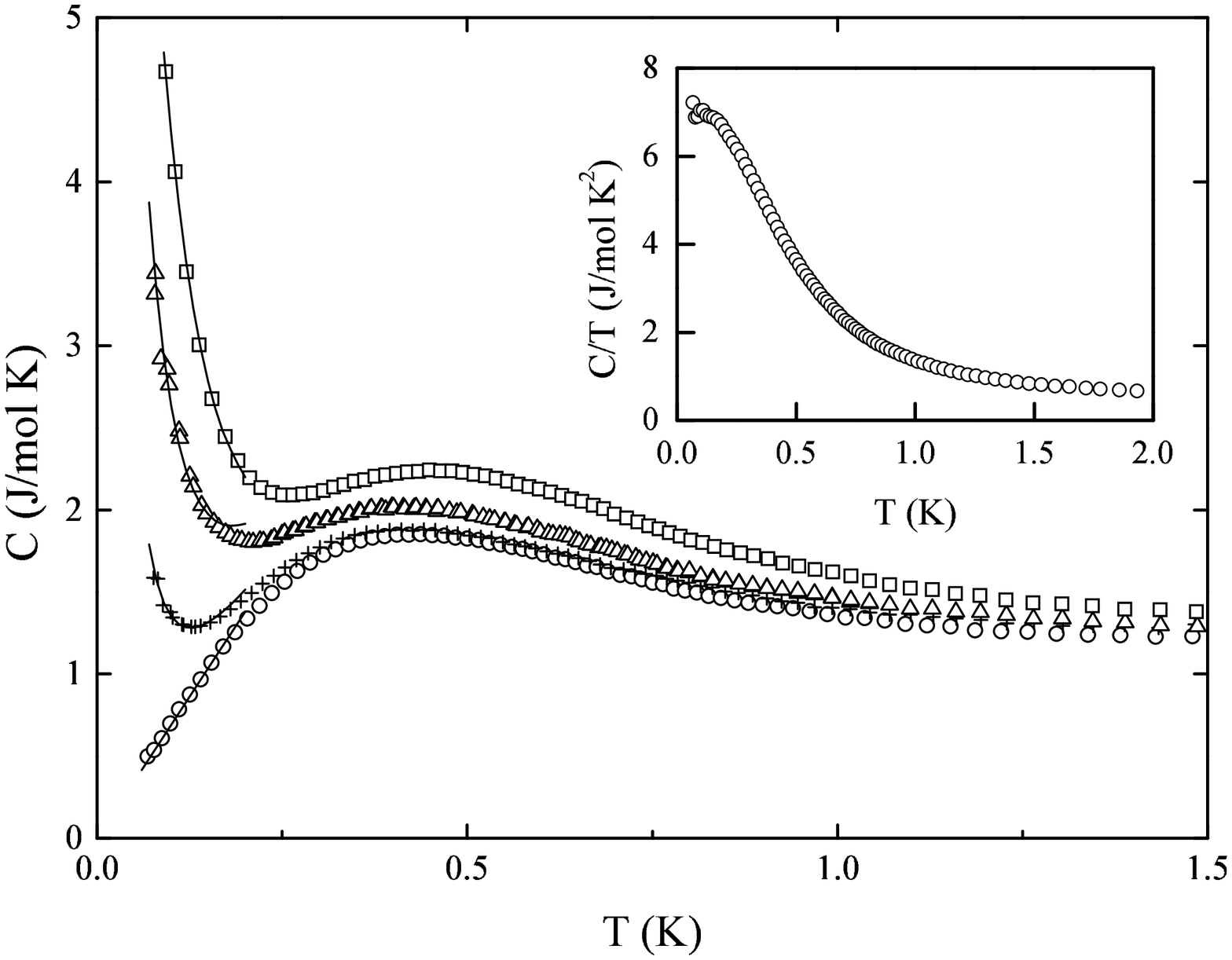}}
\caption{Specific heat of PrInAg$\rm_2$ in magnetic field. 
 ($\circ$) $H = 0$ T; (+) $H = 3$ T; ($\triangle$) $H = 6$ T; ($\Box$) $H = 9$ T. Solid lines are results of fit of the data for $T < 0.2$ K to equation~(\ref{c_Schottky}). Inset: Specific heat divided by temperature ($\gamma$) for $H = 0$ T. $\gamma (T \rightarrow 0) \approx 7$ $\rm J/mol\ K^2$. }
\label{hc_total}
\end{figure}

\noindent field-free region of the apparatus. We used a PrInAg$\rm_2$ sample with a mass of $\approx 30$ mg for the heat capacity measurements.

Fig.~\ref{hc_total} shows specific heat data of PrInAg$\rm_2$ for several values of the magnetic field up to 9 T. The inset of Fig.~\ref{hc_total} shows specific heat divided by temperature as a function of temperature in zero field. The value of the Sommerfeld coefficient (as $T \rightarrow 0$) is $\gamma \approx 7$ $\rm J/mol\ K^2$, which corresponds to a Kondo temperature $T_K = 1.29 \pi R / 6 \gamma \approx 0.8$ K within a single-channel Kondo impurity model.~\cite{rajan83}  We can make two observations about the changes in the data displayed in the main body of Fig.~\ref{hc_total} as the field is increased from zero. 

First, a low temperature tail with approximately $1/T^2$ behavior appears where the magnetic field is non-zero, and it grows as the field is increased. Such behavior is expected for a magnetic Schottky anomaly from the nuclear moments of the Pr ions with a non-magnetic electronic ground state in \prinag. The magnetic field at the nuclear site can be dramatically enhanced by the hyperfine interaction when the conduction electrons are polarized by the applied field. For \prinag \ this polarization is possible because of a Van Vleck susceptibility between the non-magnetic ground state and a higher-lying CEF split magnetic level. The 6-fold degeneracy of the nuclear magnetic manifold (nuclear spin $I = 5/2$ for Pr) is lifted by hyperfine (and external) field with an energy splitting 

\begin{equation}
\Delta = g_N H (1 + K_{hf}),
\label{delta}
\end{equation}

where $g_N = 13$ MHz/T is the nuclear gyromagnetic coefficient for Pr, and $H$ is the  applied field.   The hyperfine enhancement factor $K_{hf}$ is proportional to the magnetic susceptibility $\chi$, where the proportionality constant depends only on the properties  of the f-atom itself.~\cite{bleany73} Using

\begin{figure}
\epsfxsize=3.5in
\centerline{\epsfbox{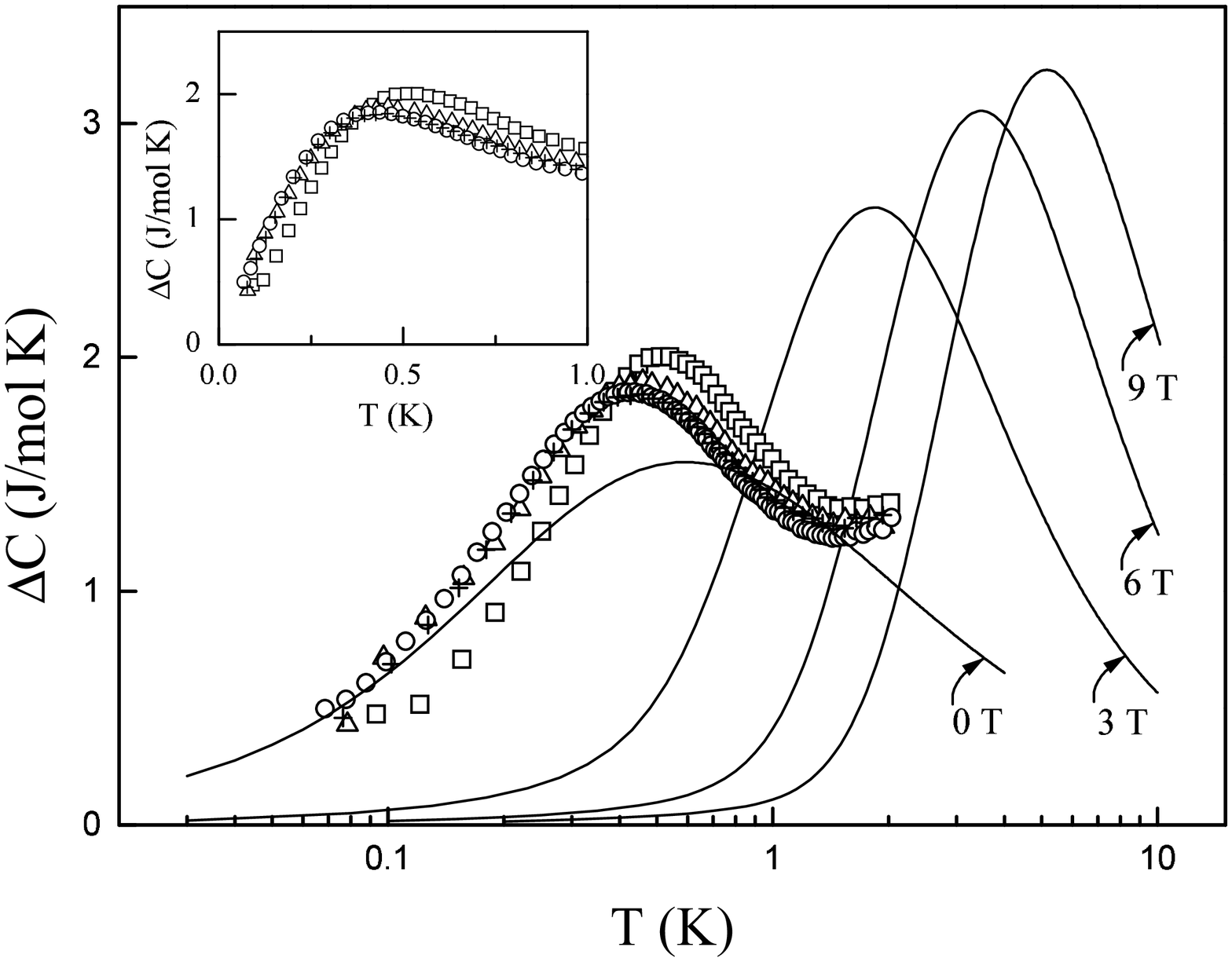}}
\caption{Specific heat of PrInAg$\rm_2$ in magnetic field after subtraction of magnetic Schottky anomaly as described in the text. ($\circ$) $H = 0$ T; (+) $H = 3$ T; ($\triangle$) $H = 6$ T; ($\Box$) $H = 9$ T. Solid lines are the results of the resonant-level model calculations~\cite{schotte75} with $J = 1/2$.  Inset: Specific heat data (low temperature Schottky anomaly subtracted) for $T < 1$ K on a linear scale.}
\label{hc_quad}
\end{figure}

\noindent    the measured value of the magnetic susceptibility at low temperatures $\chi \sim 0.04 $ emu/mol,~\cite{yatskar96} the estimated enhancement factor is $K_{hf} \approx 7.6.$~\cite{bleany73} The hyperfine enhanced splitting $\Delta$ is then determined to be 0.016 K, 0.032 K, and 0.048 K for the field of 3 T, 6 T, and 9 T, respectively. This will appear in the specific heat as a Schottky anomaly.

To compare these estimates with our experimental results, the data below 200 mK in Fig.~\ref{hc_total} was fit (the solid lines) with the expression

\begin{equation}
C(T) = \gamma T + C_{Sch}(T,\Delta ),
\label{c_Schottky}
\end{equation}

\noindent where $\gamma T$ is the electronic contribution, and $C_{Sch}$ represents the nuclear magnetic Schottky anomaly~\cite{gopal66} for Pr. The commonly used high temperature approximation $C_{Sch}(T) \propto 1/T^2$ is not adequate since the energy gap $\Delta$ is significant in comparison with the temperature range of the data. Therefore the full functional dependence of $C_{sch}$ was used in fitting. In the fit, $\gamma$ and $\Delta$ are adjustable parameters, and the resulting values of $\Delta$ are 0.017 K, 0.031 K and 0.048 K at 3 T, 6 T, and 9 T, which is in very good agreement with the values estimated above from the experimentally determined $\chi$. On the basis of this analysis we can firmly identify the low temperature tail in specific heat as a hyperfine enhanced nuclear Schottky anomaly of the Pr ions. The consistency between the value taken for the low temperature susceptibility and the low temperature specific heat in magnetic field in turn provides strong support for identification of the magnetic susceptibility as Van Vleck in nature.

The values of $\gamma$ obtained by fitting Eq.~(\ref{c_Schottky}) to the data are 6.7, 6.8, and 4.7 $\rm J/mol\ K^2$ for field of 3, 6, and 9 T respectively, with estimated uncertainty in $\gamma$ on the order of 10\%. This leads us to a second observation, that the zero-field anomaly, peaked at $\sim 420$ mK for $H = 0$~T, does not appear to be significantly affected by applied magnetic fields up to 9 T, which has corresponding energy $\mu_B H \approx 6$ K, an order of magnitude greater than the energy scale ($T_K = 0.8\rm$ K) of the feature.  This is demonstrated better in Fig.~\ref{hc_quad}, where the data are plotted on a logarithmic temperature scale for different fields after subtracting the Schottky anomaly $C_{Sch}(T, \Delta )$, determined by the fit to Eq.~(\ref{c_Schottky}). The data for H = 0 T and 3 T practically lie on top of each other, with the peak of anomaly at 0.42 K. For H = 6 T, the peak shifts slightly up to 0.45 K, and it reaches 0.5 K when the field is 9 T, which is less than a 20\% increase.

This weak dependence on the magnetic field very strongly supports the non-magnetic nature of the low-temperature specific heat anomaly associated with electron-electron interactions in PrInAg$\rm_2$. To emphasize this point, the results of the numerical calculation of a resonant-level single (magnetic) impurity Kondo model~\cite{schotte75} with $J = 1/2$ and $T_K = 0.8$ K are also plotted as the solid lines in Fig.~\ref{hc_quad}. As expected, a field of 9 T completely suppresses specific heat due to Kondo effect in the experimentally measured temperature range  and moves the anomaly almost an order of magnitude higher in temperature. A significant increase in height is also realized in the calculation as the Kondo peak narrows into a Schottky anomaly from an uncorrelated electron gas in a magnetic field. Clearly, the observed field dependence of the specific heat in PrInAg$\rm_2$ is completely different from that expected for an interaction involving the electron spin. 

The dependence of the 0.4 K specific heat anomaly on magnetic field is weak, but clearly non-zero. The change in $\gamma$ is an order magnitude greater than that in \ube13.~\cite{stewart84} One possible explanation is provided by the model that involves the splitting of the ground state $\Gamma_3$ doublet.~\cite{cox_private} This splitting $\Delta_0$ results from dynamical lifting of the orbital degeneracy of the  conduction electron bands at the Pr sites, and leads to a crossover from the two-channel quadrupolar Kondo behavior above the  temperature of $\Delta_0^2/T_K$ to the Fermi-liquid behavior below it. The additional splitting due to magnetic field is proportional to $H^4$, and the large power-law exponent results in a sharp increase of the effect of the magnetic field between 6 and 9 T. This model can quantitatively explain the observed shifts of the peak in the specific heat for 6 T and 9 T data to within a factor of two.~\cite{cox_private}

In conclusion, the specific heat of \prinag \ in a magnetic field provides two additional facts in evidence of the novel heavy-electron ground state in this compound, which  develops as a result of a Kondo-like interaction between the conduction electrons and the fluctuating quadrupolar moment of the Pr ions in a non-Kramers $\Gamma \rm _3$ ground state. The first is that the low temperature anomaly in specific heat of PrInAg$\rm_2$ (with $\gamma \approx 7$ $\rm J/mol\ K^2$) remains largely independent of magnetic field up to 9 T. The second fact is that the low temperature tail developing at non-zero field is due to a hyperfine enhanced nuclear Schottky anomaly of Pr ions. This enhancement results from the temperature independent Van Vleck susceptibility of the Pr$^{3+}$ ions in a non-magnetic ground state. The magnitude of the enhancement confirms that all of the measured low-temperature magnetic susceptibility $\chi$ is due to Pr ions, in accord with the calculations of $\chi$ based on a CEF scheme with the non-magnetic non-Kramers $\Gamma \rm _3$ ground state.~\cite{yatskar96} 

We acknowledge Y. Takano for bringing to our attention the possibility of the hyperfine enhanced nuclear Schottky origin of the low temperature tail in \prinag, and D. MacLaughlin and D. L. Cox for stimulating discussions. Work at Los Alamos and Ames National Laboratories was performed under the auspices  of the Department of Energy. Work at UC Riverside was supported by the National Science Foundation under Grant No. DMR-9624778.


\end{document}